\documentclass[]{aastex631}

\usepackage{xspace}
\usepackage{natbib}
\usepackage{multirow}

\def\arcsec{{\rm\thinspace arcsec}}

\def\g{{\rm\thinspace g}}

\def\G{{\rm\thinspace G}}

\def\K{{\rm\thinspace K}}
\def\keV{{\rm\thinspace keV}}

\def\m{{\rm\thinspace m}}

\def\Msun{\hbox{$\rm\thinspace M_{\odot}$}}

\def\s{{\rm\thinspace s}}

\def\arcsec{\hbox{$^{\prime\prime}$}}
\def\deg{\hbox{$^\circ$}}
\def\sun{\hbox{$\odot$}}
\def\earth{\hbox{$\oplus$}}
\def\lae{\mathrel{\raise .4ex\hbox{\rlap{$<$}\lower 1.2ex\hbox{$\sim$}}}}
\def\gae{\mathrel{\raise .4ex\hbox{\rlap{$>$}\lower 1.2ex\hbox{$\sim$}}}}

\newcommand{\chandra}{{Chandra}\xspace}

\newcommand{\ixpe}{{IXPE}\xspace}
\newcommand{\nicer}{{NICER}\xspace}

\newcommand{\Rm}[1]{\uppercase\expandafter{\romannumeral #1\relax}}

\newcommand{\nex}{\ion{Ne}{10}\xspace}
\newcommand{\neix}{\ion{Ne}{9}\xspace}

\newcommand{\oviii}{\ion{O}{8}\xspace}
\newcommand{\ovii}{\ion{O}{7}\xspace}

\newcommand{\fouru}{4U~1626\ensuremath{-}67\xspace}

\accepted{\today}

\submitjournal{Astrophysical Journal}

\shorttitle{Observations of 4U 1626$-$67 with the Imaging X-ray Polarimetry Explorer}
\shortauthors{Marshall et al.}

\begin{document}

\title{Observations of 4U 1626$-$67 with the Imaging X-ray Polarimetry Explorer}


\correspondingauthor{Herman L.\ Marshall}
\email{hermanm@mit.edu}

\author[0000-0002-6492-1293]{Herman L. Marshall}
\affiliation{MIT Kavli Institute for Astrophysics and Space Research, Massachusetts Institute of Technology, 77 Massachusetts Avenue, Cambridge, MA 02139, USA}
\author[0000-0002-0940-6563]{Mason Ng}
\affiliation{MIT Kavli Institute for Astrophysics and Space Research, Massachusetts Institute of Technology, 77 Massachusetts Avenue, Cambridge, MA 02139, USA}
\author[0000-0002-5359-9497]{Daniele Rogantini}
\affiliation{MIT Kavli Institute for Astrophysics and Space Research, Massachusetts Institute of Technology, 77 Massachusetts Avenue, Cambridge, MA 02139, USA}
\author[0000-0001-9739-367X]{Jeremy Heyl}
\affiliation{University of British Columbia, Vancouver, BC V6T 1Z4, Canada}
\author[0000-0002-9679-0793]{Sergey S.\ Tsygankov}
\affiliation{Department of Physics and Astronomy, 20014 University of Turku, Finland}
\affiliation{Space Research Institute of the Russian Academy of Sciences, Profsoyuznaya Str. 84/32, Moscow 117997, Russia}
\author[0000-0002-0983-0049]{Juri Poutanen}
\affiliation{Department of Physics and Astronomy, 20014 University of Turku, Finland}
\affiliation{Space Research Institute of the Russian Academy of Sciences, Profsoyuznaya Str. 84/32, Moscow 117997, Russia}
\author[0000-0003-4925-8523]{Enrico Costa}
\affiliation{INAF Istituto di Astrofisica e Planetologia Spaziali, Via del Fosso del Cavaliere 100, 00133 Roma, Italy}
\author[0000-0001-5326-880X]{Silvia Zane}
\affiliation{Mullard Space Science Laboratory, University College London, Holmbury St Mary, Dorking, Surrey RH5 6NT, UK}
\author[0000-0002-0380-0041]{Christian Malacaria}
\affiliation{International Space Science Institute, Hallerstrasse 6, 3012 Bern, Switzerland}
\author[0000-0002-3777-6182]{Iv\'an Agudo}
\affiliation{Instituto de Astrof\'{i}sicade Andaluc\'{i}a, IAA-CSIC, Glorieta de la Astronom\'{i}a s/n, 18008 Granada, Spain}
\author[0000-0002-5037-9034]{Lucio A. Antonelli}
\affiliation{INAF Osservatorio Astronomico di Roma, Via Frascati 33, 00078 Monte Porzio Catone (RM), Italy}
\affiliation{Space Science Data Center, Agenzia Spaziale Italiana, Via del Politecnico snc, 00133 Roma, Italy}
\author[0000-0002-4576-9337]{Matteo Bachetti}
\affiliation{INAF Osservatorio Astronomico di Cagliari, Via della Scienza 5, 09047 Selargius (CA), Italy}
\author[0000-0002-9785-7726]{Luca Baldini}
\affiliation{Istituto Nazionale di Fisica Nucleare, Sezione di Pisa, Largo B. Pontecorvo 3, 56127 Pisa, Italy}
\affiliation{Dipartimento di Fisica, Universit\`a di Pisa, Largo B. Pontecorvo 3, 56127 Pisa, Italy}
\author[0000-0002-5106-0463]{Wayne H. Baumgartner}
\affiliation{NASA Marshall Space Flight Center, Huntsville, AL 35812, USA}
\author[0000-0002-2469-7063]{Ronaldo Bellazzini}
\affiliation{Istituto Nazionale di Fisica Nucleare, Sezione di Pisa, Largo B. Pontecorvo 3, 56127 Pisa, Italy}
\author[0000-0002-4622-4240]{Stefano Bianchi}
\affiliation{Dipartimento di Matematica e Fisica, Universit\`a degli Studi Roma Tre, Via della Vasca Navale 84, 00146 Roma, Italy}
\author[0000-0002-0901-2097]{Stephen D. Bongiorno}
\affiliation{NASA Marshall Space Flight Center, Huntsville, AL 35812, USA}
\author[0000-0002-4264-1215]{Raffaella Bonino}
\affiliation{Istituto Nazionale di Fisica Nucleare, Sezione di Torino, Via Pietro Giuria 1, 10125 Torino, Italy}
\affiliation{Dipartimento di Fisica, Universit\`a degli Studi di Torino, Via Pietro Giuria 1, 10125 Torino, Italy}
\author[0000-0002-9460-1821]{Alessandro Brez}
\affiliation{Istituto Nazionale di Fisica Nucleare, Sezione di Pisa, Largo B. Pontecorvo 3, 56127 Pisa, Italy}
\author[0000-0002-8848-1392]{Niccol\`o Bucciantini}
\affiliation{INAF Osservatorio Astrofisico di Arcetri, Largo Enrico Fermi 5, 50125 Firenze, Italy}
\affiliation{Dipartimento di Fisica e Astronomia, Universit\`a degli Studi di Firenze, Via Sansone 1, 50019 Sesto Fiorentino (FI), Italy}
\affiliation{Istituto Nazionale di Fisica Nucleare, Sezione di Firenze, Via Sansone 1, 50019 Sesto Fiorentino (FI), Italy}
\author[0000-0002-6384-3027]{Fiamma Capitanio}
\affiliation{INAF Istituto di Astrofisica e Planetologia Spaziali, Via del Fosso del Cavaliere 100, 00133 Roma, Italy}
\author[0000-0003-1111-4292]{Simone Castellano}
\affiliation{Istituto Nazionale di Fisica Nucleare, Sezione di Pisa, Largo B. Pontecorvo 3, 56127 Pisa, Italy}
\author[0000-0001-7150-9638]{Elisabetta Cavazzuti}
\affiliation{Agenzia Spaziale Italiana, Via del Politecnico snc, 00133 Roma, Italy}
\author[0000-0002-0712-2479]{Stefano Ciprini}
\affiliation{Istituto Nazionale di Fisica Nucleare, Sezione di Roma ``Tor Vergata'', Via della Ricerca Scientifica 1, 00133 Roma, Italy}
\affiliation{Space Science Data Center, Agenzia Spaziale Italiana, Via del Politecnico snc, 00133 Roma, Italy}
\author[0000-0001-5668-6863]{Alessandra De Rosa}
\affiliation{INAF Istituto di Astrofisica e Planetologia Spaziali, Via del Fosso del Cavaliere 100, 00133 Roma, Italy}
\author[0000-0002-3013-6334]{Ettore Del Monte}
\affiliation{INAF Istituto di Astrofisica e Planetologia Spaziali, Via del Fosso del Cavaliere 100, 00133 Roma, Italy}
\author[0000-0002-5614-5028]{Laura Di Gesu}
\affiliation{Agenzia Spaziale Italiana, Via del Politecnico snc, 00133 Roma, Italy}
\author[0000-0002-7574-1298]{Niccol\`o Di Lalla}
\affiliation{Department of Physics and Kavli Institute for Particle Astrophysics and Cosmology, Stanford University, Stanford, California 94305, USA}
\author[0000-0003-0331-3259]{Alessandro Di Marco}
\affiliation{INAF Istituto di Astrofisica e Planetologia Spaziali, Via del Fosso del Cavaliere 100, 00133 Roma, Italy}
\author[0000-0002-4700-4549]{Immacolata Donnarumma}
\affiliation{Agenzia Spaziale Italiana, Via del Politecnico snc, 00133 Roma, Italy}
\author[0000-0001-8162-1105]{Victor Doroshenko}
\affiliation{Institut f\"ur Astronomie und Astrophysik, Universit\"at T\"ubingen, Sand 1, 72076 T\"ubingen, Germany}
\author[0000-0003-0079-1239]{Michal Dov\v{c}iak}
\affiliation{Astronomical Institute of the Czech Academy of Sciences, Bo\v{c}n\'i II 1401/1, 14100 Praha 4, Czech Republic}
\author[0000-0003-4420-2838]{Steven R. Ehlert}
\affiliation{NASA Marshall Space Flight Center, Huntsville, AL 35812, USA}
\author[0000-0003-1244-3100]{Teruaki Enoto}
\affiliation{RIKEN Cluster for Pioneering Research, 2-1 Hirosawa, Wako, Saitama 351-0198, Japan}
\author[0000-0001-6096-6710]{Yuri Evangelista}
\affiliation{INAF Istituto di Astrofisica e Planetologia Spaziali, Via del Fosso del Cavaliere 100, 00133 Roma, Italy}
\author[0000-0003-1533-0283]{Sergio Fabiani}
\affiliation{INAF Istituto di Astrofisica e Planetologia Spaziali, Via del Fosso del Cavaliere 100, 00133 Roma, Italy}
\author[0000-0003-1074-8605]{Riccardo Ferrazzoli}
\affiliation{INAF Istituto di Astrofisica e Planetologia Spaziali, Via del Fosso del Cavaliere 100, 00133 Roma, Italy}
\author[0000-0003-3828-2448]{Javier A. Garcia}
\affiliation{California Institute of Technology, Pasadena, CA 91125, USA}
\author[0000-0002-5881-2445]{Shuichi Gunji}
\affiliation{Yamagata University,1-4-12 Kojirakawa-machi, Yamagata-shi 990-8560, Japan}
\author{Kiyoshi Hayashida}
\affiliation{Osaka University, 1-1 Yamadaoka, Suita, Osaka 565-0871, Japan}
\author[0000-0002-0207-9010]{Wataru Iwakiri}
\affiliation{Department of Physics, Faculty of Science and Engineering, Chuo University, 1-13-27 Kasuga, Bunkyo-ku, Tokyo 112-8551, Japan}
\author[0000-0001-9522-5453]{Svetlana G. Jorstad}
\affiliation{Institute for Astrophysical Research, Boston University, 725 Commonwealth Avenue, Boston, MA 02215, USA}
\affiliation{Department of Astrophysics, St. Petersburg State University, Universitetsky pr. 28, Petrodvoretz, 198504 St. Petersburg, Russia}
\author[0000-0002-5760-0459]{Vladimir Karas}
\affiliation{Astronomical Institute of the Czech Academy of Sciences, Bo\v{c}n\'i II 1401/1, 14100 Praha 4, Czech Republic}
\author{Takao Kitaguchi}
\affiliation{RIKEN Cluster for Pioneering Research, 2-1 Hirosawa, Wako, Saitama 351-0198, Japan}
\author[0000-0002-0110-6136]{Jeffery J. Kolodziejczak}
\affiliation{NASA Marshall Space Flight Center, Huntsville, AL 35812, USA}
\author[0000-0002-1084-6507]{Henric Krawczynski}
\affiliation{Physics Department and McDonnell Center for the Space Sciences, Washington University in St. Louis, St. Louis, MO 63130, USA}
\author[0000-0001-8916-4156]{Fabio La Monaca}
\affiliation{INAF Istituto di Astrofisica e Planetologia Spaziali, Via del Fosso del Cavaliere 100, 00133 Roma, Italy}
\author[0000-0002-0984-1856]{Luca Latronico}
\affiliation{Istituto Nazionale di Fisica Nucleare, Sezione di Torino, Via Pietro Giuria 1, 10125 Torino, Italy}
\author[0000-0001-9200-4006]{Ioannis Liodakis}
\affiliation{Finnish Centre for Astronomy with ESO, 20014 University of Turku, Finland}
\author[0000-0002-0698-4421]{Simone Maldera}
\affiliation{Istituto Nazionale di Fisica Nucleare, Sezione di Torino, Via Pietro Giuria 1, 10125 Torino, Italy}
\author[0000-0002-0998-4953]{Alberto Manfreda}
\affiliation{Istituto Nazionale di Fisica Nucleare, Sezione di Pisa, Largo B. Pontecorvo 3, 56127 Pisa, Italy}
\author[0000-0003-4952-0835]{Fr\'{e}d\'{e}ric Marin}
\affiliation{Universit\'{e} de Strasbourg, CNRS, Observatoire Astronomique de Strasbourg, UMR 7550, 67000 Strasbourg, France}
\author[0000-0002-2055-4946]{Andrea Marinucci}
\affiliation{Agenzia Spaziale Italiana, Via del Politecnico snc, 00133 Roma, Italy}
\author[0000-0001-7396-3332]{Alan P. Marscher}
\affiliation{Institute for Astrophysical Research, Boston University, 725 Commonwealth Avenue, Boston, MA 02215, USA}
\author[0000-0002-2152-0916]{Giorgio Matt}
\affiliation{Dipartimento di Matematica e Fisica, Universit\`a degli Studi Roma Tre, Via della Vasca Navale 84, 00146 Roma, Italy}
\author{Ikuyuki Mitsuishi}
\affiliation{Graduate School of Science, Division of Particle and Astrophysical Science, Nagoya University, Furo-cho, Chikusa-ku, Nagoya, Aichi 464-8602, Japan}
\author[0000-0001-7263-0296]{Tsunefumi Mizuno}
\affiliation{Hiroshima Astrophysical Science Center, Hiroshima University, 1-3-1 Kagamiyama, Higashi-Hiroshima, Hiroshima 739-8526, Japan}
\author[0000-0003-3331-3794]{Fabio Muleri}
\affiliation{INAF Istituto di Astrofisica e Planetologia Spaziali, Via del Fosso del Cavaliere 100, 00133 Roma, Italy}
\author[0000-0002-5847-2612]{C.-Y. Ng}
\affiliation{Department of Physics, The University of Hong Kong, Pokfulam, Hong Kong}
\author[0000-0002-1868-8056]{Stephen L. O'Dell}
\affiliation{NASA Marshall Space Flight Center, Huntsville, AL 35812, USA}
\author[0000-0002-5448-7577]{Nicola Omodei}
\affiliation{Department of Physics and Kavli Institute for Particle Astrophysics and Cosmology, Stanford University, Stanford, California 94305, USA}
\author[0000-0001-6194-4601]{Chiara Oppedisano}
\affiliation{Istituto Nazionale di Fisica Nucleare, Sezione di Torino, Via Pietro Giuria 1, 10125 Torino, Italy}
\author[0000-0001-6289-7413]{Alessandro Papitto}
\affiliation{INAF Osservatorio Astronomico di Roma, Via Frascati 33, 00078 Monte Porzio Catone (RM), Italy}
\author[0000-0002-7481-5259]{George G. Pavlov}
\affiliation{Department of Astronomy and Astrophysics, Pennsylvania State University, University Park, PA 16802, USA}
\author[0000-0001-6292-1911]{Abel L. Peirson}
\affiliation{Department of Physics and Kavli Institute for Particle Astrophysics and Cosmology, Stanford University, Stanford, California 94305, USA}
\author[0000-0003-3613-4409]{Matteo Perri}
\affiliation{Space Science Data Center, Agenzia Spaziale Italiana, Via del Politecnico snc, 00133 Roma, Italy}
\affiliation{INAF Osservatorio Astronomico di Roma, Via Frascati 33, 00078 Monte Porzio Catone (RM), Italy}
\author[0000-0003-1790-8018]{Melissa Pesce-Rollins}
\affiliation{Istituto Nazionale di Fisica Nucleare, Sezione di Pisa, Largo B. Pontecorvo 3, 56127 Pisa, Italy}
\author[0000-0001-6061-3480]{Pierre-Olivier Petrucci}
\affiliation{Universit\'e Grenoble Alpes, CNRS, IPAG, 38000 Grenoble, France}
\author[0000-0001-7397-8091]{Maura Pilia}
\affiliation{INAF Osservatorio Astronomico di Cagliari, Via della Scienza 5, 09047 Selargius (CA), Italy}
\author[0000-0001-5902-3731]{Andrea Possenti}
\affiliation{INAF Osservatorio Astronomico di Cagliari, Via della Scienza 5, 09047 Selargius (CA), Italy}
\author[0000-0002-2734-7835]{Simonetta Puccetti}
\affiliation{Space Science Data Center, Agenzia Spaziale Italiana, Via del Politecnico snc, 00133 Roma, Italy}
\author[0000-0003-1548-1524]{Brian D. Ramsey}
\affiliation{NASA Marshall Space Flight Center, Huntsville, AL 35812, USA}
\author[0000-0002-9774-0560]{John Rankin}
\affiliation{INAF Istituto di Astrofisica e Planetologia Spaziali, Via del Fosso del Cavaliere 100, 00133 Roma, Italy}
\author[0000-0003-0411-4243]{Ajay Ratheesh}
\affiliation{INAF Istituto di Astrofisica e Planetologia Spaziali, Via del Fosso del Cavaliere 100, 00133 Roma, Italy}
\author[0000-0001-6711-3286]{Roger W. Romani}
\affiliation{Department of Physics and Kavli Institute for Particle Astrophysics and Cosmology, Stanford University, Stanford, California 94305, USA}
\author[0000-0001-5676-6214]{Carmelo Sgr\`o}
\affiliation{Istituto Nazionale di Fisica Nucleare, Sezione di Pisa, Largo B. Pontecorvo 3, 56127 Pisa, Italy}
\author[0000-0002-6986-6756]{Patrick Slane}
\affiliation{Harvard \& Smithsonian Center for Astrophysics, 60 Garden St, Cambridge, MA 02138, USA}
\author[0000-0001-8916-4156]{Paolo Soffitta}
\affiliation{INAF Istituto di Astrofisica e Planetologia Spaziali, Via del Fosso del Cavaliere 100, 00133 Roma, Italy}
\author[0000-0003-0802-3453]{Gloria Spandre}
\affiliation{Istituto Nazionale di Fisica Nucleare, Sezione di Pisa, Largo B. Pontecorvo 3, 56127 Pisa, Italy}
\author[0000-0002-8801-6263]{Toru Tamagawa}
\affiliation{RIKEN Cluster for Pioneering Research, 2-1 Hirosawa, Wako, Saitama 351-0198, Japan}
\author[0000-0003-0256-0995]{Fabrizio Tavecchio}
\affiliation{INAF Osservatorio Astronomico di Brera, Via E. Bianchi 46, 23807 Merate (LC), Italy}
\author[0000-0002-1768-618X]{Roberto Taverna}
\affiliation{Dipartimento di Fisica e Astronomia, Universit\`a degli Studi di Padova, Via Marzolo 8, 35131 Padova, Italy}
\author{Yuzuru Tawara}
\affiliation{Graduate School of Science, Division of Particle and Astrophysical Science, Nagoya University, Furo-cho, Chikusa-ku, Nagoya, Aichi 464-8602, Japan}
\author[0000-0002-9443-6774]{Allyn F. Tennant}
\affiliation{NASA Marshall Space Flight Center, Huntsville, AL 35812, USA}
\author[0000-0003-0411-4606]{Nicholas E. Thomas}
\affiliation{NASA Marshall Space Flight Center, Huntsville, AL 35812, USA}
\author[0000-0002-6562-8654]{Francesco Tombesi}
\affiliation{Dipartimento di Fisica, Universit\`a degli Studi di Roma ``Tor Vergata'', Via della Ricerca Scientifica 1, 00133 Roma, Italy}
\affiliation{Istituto Nazionale di Fisica Nucleare, Sezione di Roma ``Tor Vergata'', Via della Ricerca Scientifica 1, 00133 Roma, Italy}
\affiliation{Department of Astronomy, University of Maryland, College Park, Maryland 20742, USA}
\author[0000-0002-3180-6002]{Alessio Trois}
\affiliation{INAF Osservatorio Astronomico di Cagliari, Via della Scienza 5, 09047 Selargius (CA), Italy}
\author[0000-0003-3977-8760]{Roberto Turolla}
\affiliation{Dipartimento di Fisica e Astronomia, Universit\`a degli Studi di Padova, Via Marzolo 8, 35131 Padova, Italy}
\affiliation{Mullard Space Science Laboratory, University College London, Holmbury St Mary, Dorking, Surrey RH5 6NT, UK}
\author[0000-0002-4708-4219]{Jacco Vink}
\affiliation{Anton Pannekoek Institute for Astronomy \& GRAPPA, University of Amsterdam, Science Park 904, 1098 XH Amsterdam, The Netherlands}
\author[0000-0002-5270-4240]{Martin C. Weisskopf}
\affiliation{NASA Marshall Space Flight Center, Huntsville, AL 35812, USA}
\author[0000-0002-7568-8765]{Kinwah Wu}
\affiliation{Mullard Space Science Laboratory, University College London, Holmbury St Mary, Dorking, Surrey RH5 6NT, UK}
\author[0000-0002-0105-5826]{Fei Xie}
\affiliation{Guangxi Key Laboratory for Relativistic Astrophysics, School of Physical Science and Technology, Guangxi University, Nanning 530004, China}
\affiliation{INAF Istituto di Astrofisica e Planetologia Spaziali, Via del Fosso del Cavaliere 100, 00133 Roma, Italy}
\collaboration{92}{(IXPE Collaboration)}
\nocollaboration{2}
\author[0000-0002-0842-7792]{Norbert S.\ Schulz}
\affiliation{MIT Kavli Institute for Astrophysics and Space Research, Massachusetts Institute of Technology, 77 Massachusetts Avenue, Cambridge, MA 02139, USA}
\author[0000-0001-8804-8946]{Deepto Chakrabarty}
\affiliation{MIT Kavli Institute for Astrophysics and Space Research, Massachusetts Institute of Technology, 77 Massachusetts Avenue, Cambridge, MA 02139, USA}



\begin{abstract}
We present measurements of the polarization of X-rays in the 2--8~keV band from the pulsar in the ultracompact low mass X-ray binary \fouru\ using data from the {Imaging X-ray Polarimetry Explorer} ({IXPE}). The $7.66$ s pulsations were clearly detected throughout the \ixpe  observations as well as in the \nicer soft X-ray observations, which we use as the basis for our timing analysis and to constrain the spectral shape over 0.4--10 keV energy band.
\chandra HETGS high-resolution X-ray spectra were also obtained near the times of the \ixpe observations for firm spectral modeling.
We find an upper limit on the pulse-averaged linear polarization of $<$4\% (at 95\% confidence).  Similarly, there was no significant detection of polarized flux in pulse phase intervals when subdividing the bandpass by energy.
However, spectropolarimetric modeling over the full bandpass in pulse phase intervals provide a marginal detection of polarization of the power-law spectral component at the 4.8$\pm$2.3\% level (90\% confidence).
We discuss the implications concerning the accretion geometry onto the pulsar, favoring two-component models of the pulsed emission.
\end{abstract}


\section{Introduction} \label{sec:intro}

\fouru is an ultracompact low mass X-ray binary (UCXB), featuring a neutron star in a 42-minute orbit around a very low-mass companion, and is unique as the only UCXB hosting a persistent, strongly magnetized X-ray pulsar. The X-ray source was discovered by {Uhuru} \citep{giacconi_uhuru_1972} and was found to pulse at a period of $P =7.68$\,s using {SAS-3} data \citep[][]{rappaport_discovery_1977}. The ultracompact nature of the system had been suggested by \citet{joss_accreting_1978}, and was established by the detection of a 42-minute orbital period \citep{middleditch_4u_1981,chakrabarty_high-speed_1998}.

The pulsar steadily spun up until sometime in 1990 \citep[see][and references therein]{chakrabarty_torque_1997}, when it began a period of spin-down and dimming on a timescale of years. The pulsar exhibited a torque reversal back to spin-up in 2008 \citep{camero-arranz_new_2010}, which was accompanied by a large increase in flux \citep{camero-arranz_4u_2012}, bringing the source back to its pre-1990 brightness. The
magnetic field strength is estimated to be $B\sim (3-4)\times10^{12}$\,G, based on a cyclotron resonance scattering feature (CRSF) at $\sim$37\,keV \citep{orlandini_bepposax_1998}.  \citet{chakrabarty1997} suggested that the pulsar spins at its equilibrium period, where the Keplerian co-rotation radius, $6.5\times10^8$ cm, is comparable to the magnetospheric radius, thereby limiting the mass accretion rate to $> 2\times10^{-10}$ \Msun{} yr$^{-1}$. The early X-ray spectra of \fouru showed an emission line complex around 1\,keV, identified as \nex, along with emission from \oviii around 0.6\,keV \citep{angelini_neon_1995}. High resolution spectroscopy during the spin-down phase with \chandra showed that the hydrogenic \nex and \oviii features were double-peaked, suggesting an accretion disk origin, and identified emission from the He-like lines of \neix and \ovii \citep{schulz_doublepeaked_2001,krauss_high-resolution_2007}. 

Timing and spectroscopy has given us a wealth of information about this system, but there are open questions regarding the accretion geometry of the pulsar. With the launch of IXPE on 2021 December 9, X-ray polarization measurements can now provide 
geometric constraints useful for characterizing X-ray pulsars
\citep{meszaros88}. For example, X-ray polarization may enable us to distinguish between the fan beam (emission perpendicular to the neutron star magnetic axis) and the pencil beam (emission parallel to magnetic axis) 
{emission patterns} 
\citep{meszaros85b,meszaros88,2021MNRAS.503.5193M,sokolova21,2021MNRAS.501..109C}.
{In two previous \ixpe observations, polarizations were lower than expected, showing that the two models are not clearly distinguished.  In Her X-1, there is the possibility that both emission geometries may be operating \citep{herx1_ixpe}.  In Cen X-3, complications from atmosphere modeling and surface reflection are suggested \citep{2022arXiv220902447T}.  However, in both cases, the inclination of the pulsar spin axis and the magnetic field obliquity were measurable.  In the case of Her X-1, it appears that the spin axis is not aligned to the orbital plane, which may result from free precession of the neutron star.  Thus, even when the polarization is low, geometric parameters can be derived with interesting physical implications.}

Here we report polarimetric observations of the pulsar in the ultracompact low mass X-ray binary \fouru obtained with the {\it Imaging X-ray Polarimetry Explorer} (\ixpe).
In addition to the \ixpe observations, \fouru was also observed by \nicer and \chandra in order to provide broad-band ($\sim 0.3 - 12 \keV$) measurements of the X-ray spectrum and pulse timing. We outline the observations in Section \ref{sec:observations}, present the results in Section \ref{sec:results}, and discuss the results and conclude in Section \ref{sec:discussion}.

\section{Observation Data} \label{sec:observations}

\ixpe observations of \fouru\ were taken starting on 2022 March 24 beginning at 01:51 UTC and ending on 2022 March 27 at 05:39 UTC. Nearly simultaneous observations with other X-ray telescopes along with the \ixpe observations are summarized in Table~\ref{tab:observations}. Operational constraints limited the ability of \nicer and \chandra to observe \fouru\ simultaneously with \ixpe. Given that \fouru\ has not been previously observed to exhibit large variability in the X-rays on timescales of $\sim \thinspace \mathrm{days}$, we do not anticipate this offset in time to introduce any large systematic errors in our joint spectral analysis. Light curve analysis for each of these observations also show no evidence for significant variability in \fouru's emission. 

\begin{table}[htbp]
\centering  
\begin{tabular}{cccccccc}
\hline \hline
Mission	   & ObsID & Energy    & Start date    & End date      & Exposure 	 \\
		   &       & $(\keV)$  & (YYYY-MM-DD)  & (YYYY-MM-DD)  & (ks)  	\\
\hline
\ixpe	& 01002701 & 2--8 & 2022-03-24    & 2022-03-27    & $189.5$ 	\\
\nicer & 5661010[101-103, 201-205] & 0.4--10 & 2022-03-18 & 2022-03-31 & $29.7$  \\
\chandra/HETGS & 24700 & 0.7--7.5 & 2022-01-05 & 2022-01-06 & $11.8$ \\
\chandra/HETGS & 26250 & 0.7--7.5 & 2022-01-06 & 2022-01-06 & $12.7$ \\
\chandra/HETGS & 26009 & 0.7--7.5 & 2022-04-16 & 2022-04-16 & $56.8$ \\
\chandra/HETGS & 26087 & 0.7--7.5 & 2022-04-23 & 2022-04-24 & $29.3$ \\
\hline 
\end{tabular}
\caption{Observational log of the X-ray telescope observations used in this work. \ixpe\ comprises 3 operating DUs, and their exposure times were 189.5, 189.5, and 189.3 ks, respectively. 
}\label{tab:observations}   
\end{table}

\subsection{IXPE}

\ixpe is a NASA Small Explorer mission in partnership with the Italian Space Agency (ASI). 
\ixpe includes three identical X-ray telescopes, each comprising an X-ray mirror assembly and a polarization-sensitive pixelated detector, to provide imaging polarimetry over a nominal 2--8~keV band. We refer interested readers to \cite{Weisskopf2022} for a complete description of the hardware deployed upon \ixpe and its performance, only summarizing the most relevant details of the instrument here. 

\ixpe data were processed using a pipeline that estimates the photoelectron emission direction (and hence the polarization), location, and energy of each event after applying corrections for charging effects, detector temperature, and Gas Electron Multiplier (GEM) gain non-uniformity. Spurious modulation is removed on an event-by-event basis \citep{2022AJ....163...39R}.
The output of this pipeline processing is an event file for each of the three \ixpe Detector Units (DUs) that contains, in addition to the typical information of spatially resolved X-ray astronomy, event-specific angles useful for determining the linear polarization of the radiation. 
 
Analysis of the processed \ixpe event lists is carried out with several independent analysis tools, most notably \texttt{ixpeobssim}, which is a software suite designed specifically to operate with both simulated and real \ixpe data, and \cite{Baldini2022} describes these algorithms in detail.  {In particular, we made use of the canned routines to do data filtering (\texttt{xpselect}), generate polarization cubes (\texttt{xpbin} with the \texttt{PCUBE} algorithm), and extraction polarization parameters (\texttt{xpbinview})}.  In this work, we made use of version 25.6.3 of \texttt{ixpeobssim}. {The software suite \texttt{ixpeobssim} is now publicly available on GitHub.\footnote{{https://github.com/lucabaldini/ixpeobssim}}}  Before processing the event files, we corrected for position offset and energy calibration offsets as recommended by the \ixpe team using standard \texttt{ftools} (as outlined in the README file associated with the data). 

For \fouru, the core region was filtered from the rest of the event list using a 60\arcsec{} aperture around the centroid of the X-ray data. The background region is derived from an annulus region with an inner radius of 120\arcsec{} and an outer radius of 240\arcsec. Information on the polarization in the core of \fouru\ is obtained by running the \texttt{xpbin} tool on the selected event list with the \texttt{PCUBE} algorithm. The resulting data structure, hereafter a polarization cube, measures the polarization degree and angle with their uncertainties along with minimum detectable polarization at 99\% level of significance (MDP$_{99}$) of the region in a given energy band. The polarization properties and their uncertainties are calculated using either a weighting approach \citep{2015APh....68...45K,2022AJ....163..170D} or an unbinned maximum likelihood method \citep{2021AJ....162..134M}. 

The spectropolarimetric spectra were generated using the \texttt{PHA1}, \texttt{PHA1Q}, and \texttt{PHA1U} binning algorithms as part of \texttt{xpbin}. We use the weighted version of the v10 instrument response functions \citep{2022AJ....163..170D}.
  
\subsection{NICER}

\nicer is an external payload onboard the International Space Station (ISS). It has 56 co-aligned X-ray concentrator optics and 56 focal plane modules (FPMs) containing silicon drift detectors (of which 52 are usually active). NICER has fast timing capabilities in the 0.2--12 keV band, with a timing accuracy of time-tagged photons to better than 100 ns \citep{gendreau_nicer,lamarr_nicer,prigozhin_nicer}. 

NICER observed \fouru\ for a total of 29.7 ks of filtered exposure across two intervals, over 2022 March 18 to 2022 March 20 (ObsIDs 5661010101 to 5661010103) and 2022 March 27 to 2022 March 31 (ObsIDs 5661010201 to 5661010205). The NICER observations were segmented due to visibility constraints, which happened to overlap directly with the \ixpe observations. The data from ObsID 5661010103 are unusable, due to visibility constraints likely as a result of glint off the solar panels on the ISS (priv. comm., K. Gendreau). The observations were processed using version 9 of the NICER Data Analysis Software (\texttt{NICERDAS}) in HEASOFT 6.30.1. The following filtering criteria were imposed in the construction of good time intervals (GTIs): undershoots (dark current) count rate per FPM, \texttt{underonly\_range} $<$ 500 c/s; overshoots (charged particle saturation) count rate per FPM, \texttt{overonly\_range} $<$ 1.5 c/s and \texttt{1.52$\times$1.5$\times$COR\_SAX$^{-0.633}$} c/s; $\geq$ 38 operational FPMs; pointing offset of $<0\fdg015$ from the nominal source position; $\geq 20^\circ$ for the source-Earth limb angle ($\geq 30^\circ$ for the bright Earth angle). 

The response files were generated for each ObsID using \texttt{nicerrmf} and \texttt{nicerarf} available as part of \texttt{NICERDAS}. The background spectra were generated with the \texttt{nibackgen3C50} tool \citep{remillard2022}. The spectral analysis was carried out with \texttt{XSPEC} version 12.12.1, and the spectra were rebinned to have at least 25 counts per bin, and grouped according to the optimal binning scheme \citep{kaastra2016}. 

The events were corrected for the Solar System barycenter in the ICRS reference frame, using the source coordinates R.A.$=248\fdg07$ and Decl.$=-67\fdg46092$ \citep{lin_coordinates_2012}, with the \texttt{FTOOLS barycorr} with JPL DE421 solar system ephemeris \citep{folkner_de421}.

\subsection{Chandra HETGS}

\chandra observed \fouru multiple times in January and April 2022 for a total of 24 ks (obsIDs 24700 and 26250) and 86 ks (obsIDs 26009 and 26086), respectively. All the observations were taken with the High Energy Transmission Grating Spectrometer (HETGS) in the Timed Exposure mode \citep{Canizares05}. We extracted the High Energy Grating (HEG) and Medium Energy Grating (MEG) spectra using the \texttt{CIAO} X-ray data analysis package \citep{Fruscione06}, along with the most recent calibration (CALDB) products and we reduced them following the processing procedures from the \chandra Gratings Catalog and Archive \citep[TGCat;][]{huenemoerder_tgcat:_2011}. The X-ray spectral analysis was performed using the \texttt{SPEX} fitting package \citep{Kaastra20} v3.06.01. The spectral shape of the continuum did not vary among the different observations and we combined the spectra using the {\tt combine\_grating\_spectra} tool. We fit the stacked spectrum considering the 1.7--10\,\AA\ wavelength band for HEG and 2--17\,\AA\ wavelength band for MEG. The data has been binned using the optimized binning, {\tt obin}, in \texttt{SPEX} \citep{kaastra2016}.

\section{Results} \label{sec:results}

\subsection{Pulse Phase-Averaged Polarization} \label{sec:polarization_ave}

\begin{figure}[t]
		\includegraphics[width=\linewidth]{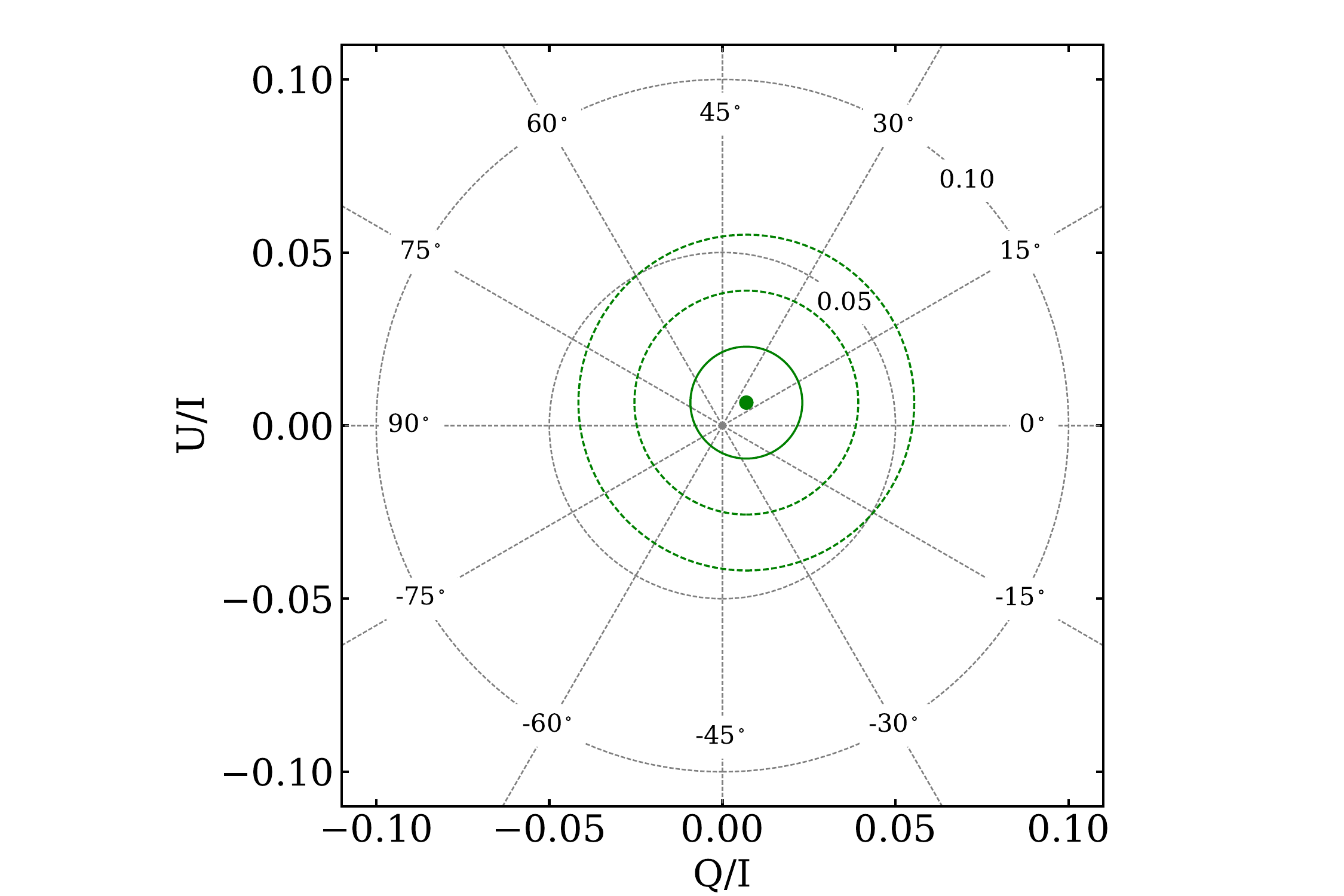}
		\caption{Pulse phase-averaged normalized Stokes parameters $U/I$ and $Q/I$ over 2.0--8.0 keV energy range. The $1\sigma, 2\sigma$, and $3\sigma$ contours are plotted as concentric circles around the best estimate. Light dotted circles are loci of constant polarization fraction for levels of 5\% and 10\%, while radial lines are labeled for specific
		electric vector position angles (EVPAs) relative to North. These results indicate that the phase-averaged polarization fraction is consistent with zero.
			\label{fig:normqu_all}}
	\end{figure}

In calculating the polarization quantities, we first followed the formalism as given in \cite{2015APh....68...45K} and \cite{Baldini2022}.
Figure~\ref{fig:normqu_all} shows the confidence contours for the normalized
Stokes parameters, $q \equiv Q/I$ and $u \equiv U/I$, for the full 2--8 keV bandpass.
Note that the 1$\sigma$ contour includes the origin, indicating that the broad-band
polarization is consistent with zero within 1$\sigma$.
The 95\% limit to the polarization fraction (the 2$\sigma$ contour in Figure~\ref{fig:normqu_all}) ranges from 2 to 4\% because it depends on the true but unknown electric vector position angle (EVPA).
We also used an unbinned maximum likelihood method \citep{2021AJ....162..134M}; the best fit values were slightly different, with $q,u$ = (0.01, $-$0.01), but the 1$\sigma$ error region also included the origin and the 2$\sigma$ confidence region for the polarization fraction ranged up to 3.7\%.  As in \citet{Ehlert2022}, an accounting for unpolarized background was included in the likelihood.

\begin{figure}[t]
		\includegraphics[width=\linewidth]{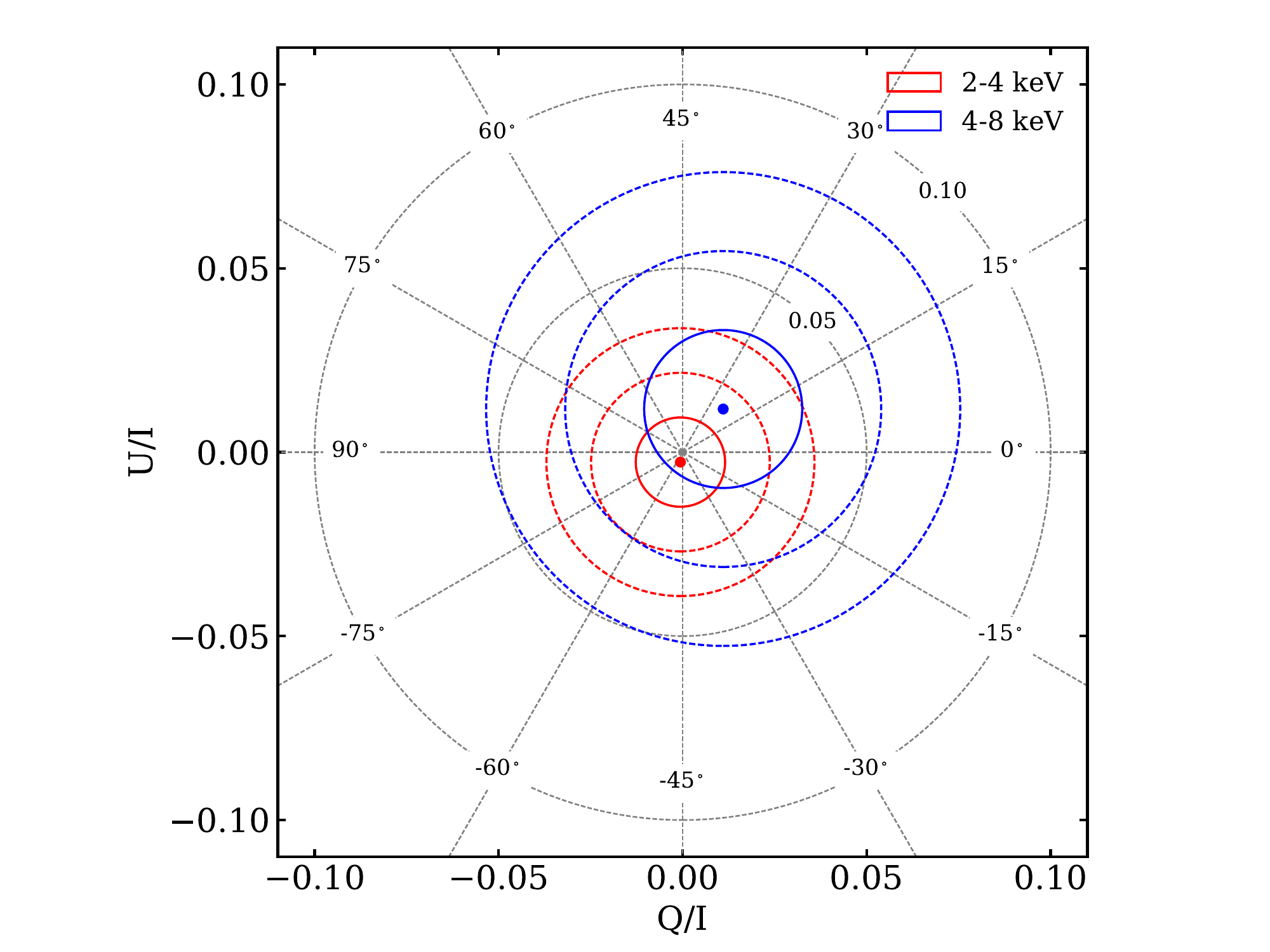}
		\caption{Same as Figure~\ref{fig:normqu_all} except for the 2.0--4.0 keV (red) and 4.0--8.0 keV (blue) bands independently.
		\label{fig:normqu}}
	\end{figure}

The data were split into two energy bands, 2--4 and 4--8 keV, to test whether
the measured polarization degree is a function of energy.  Again, as
shown in Figure~\ref{fig:normqu}, the 1$\sigma$ contours are consistent with zero polarization.
The $1\sigma$ and $3\sigma$ upper limits to the polarization
fraction are given in Table \ref{tab:pd_table}.

\begin{table}[htbp]
\centering  
\caption{Polarization estimates and upper limits in two energy bands}
\begin{tabular}{cccc}
\hline \hline
Energy band (keV) & \multicolumn{3}{c}{Polarization Degree (\%)} \\
 & Best Estimate & $1\sigma$ & $3\sigma$ \\
\hline
2.0--4.0 & $0.3$ & $<2.1$ & $<4.5$ \\
4.0--8.0 & $1.6$ & $<4.9$ & $<9.0$ \\
\hline 
\end{tabular}
\label{tab:pd_table}   
\end{table}

 	\begin{figure}[t]
		\includegraphics[width=\linewidth]{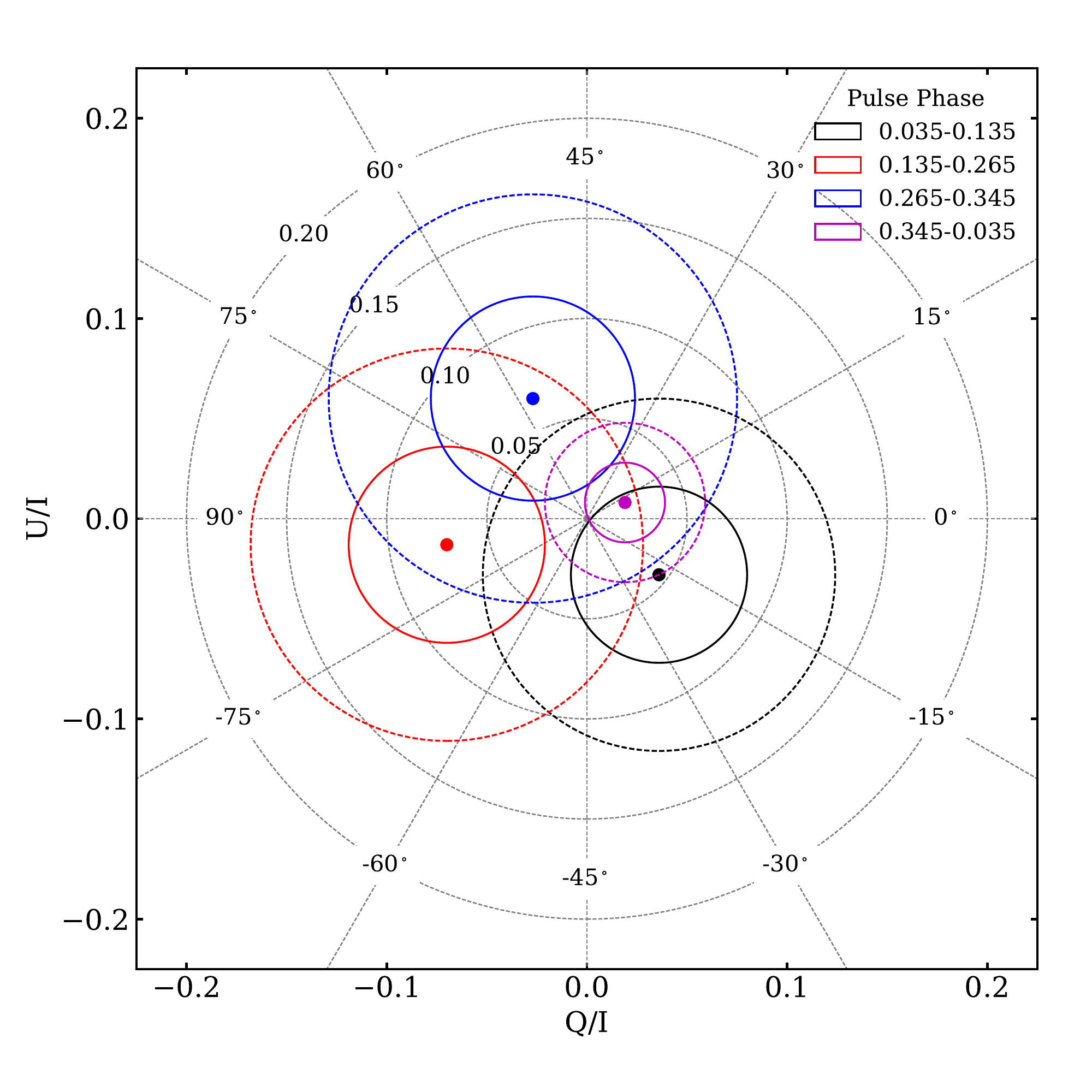}
		\caption{Normalized Stokes parameters $U/I$ and $Q/I$ as a function of pulse phase over 2.0--8.0 keV energy range (tabulated data in Table \ref{tab:qu_table}). The $1\sigma$ and $2\sigma$ contours are plotted as concentric circles around the best estimate. Light dotted circles are loci of constant polarization fraction for levels in multiples of 5\%, while radial lines are labeled for specific electric vector position angles (EVPAs) relative to North. 
		\label{fig:normqu_phase}}
	\end{figure}
 
\subsection{Timing} \label{sec:timing}

\begin{figure}[t]
		\centering
		\includegraphics[width=\linewidth]{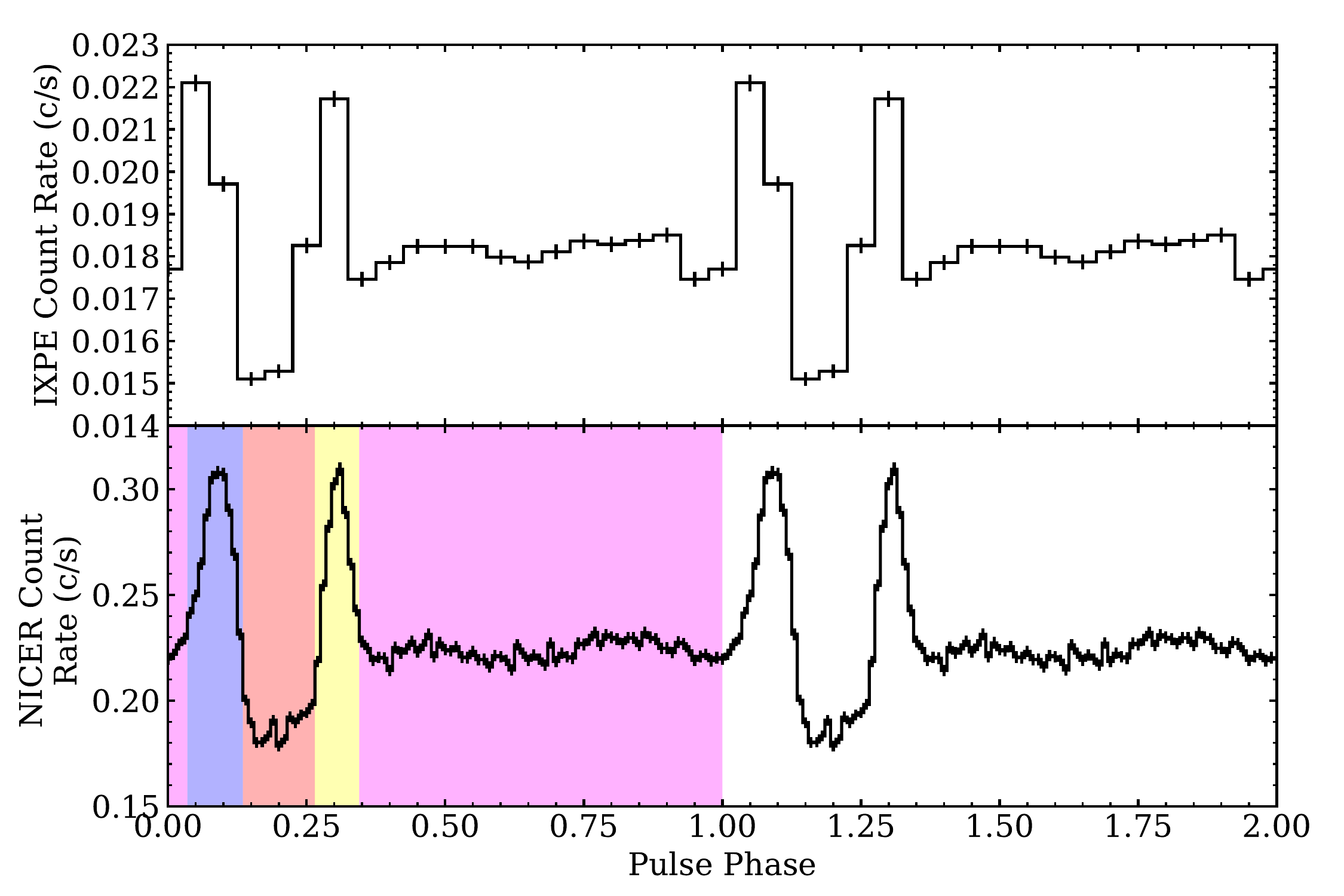}
		\caption{Top: Folded 2.0--8.0 keV pulse profile for \fouru with the combined \ixpe events from all three detectors. Bottom: Folded 2.0--8.0 keV pulse profile for \fouru using \nicer data. The two peaks correspond to pulse phase values 0.0907 and 0.3076 (derived from fitting a parabola around the peaks). The different colors represent the pulse phase intervals used in the analysis of Sect.~\ref{sec:polarization_dep}: a) 0.345--0.035 (modulo 1; pink), b) 0.035--0.135 (blue), c) 0.135--0.265 (red), and 0.265--0.345 (yellow). The phase boundaries were based on the structure of the \nicer pulse profile. The pulse profiles are phase-aligned and two rotation cycles are plotted for clarity.}
		\label{fig:pulseprofile}
	\end{figure}

In order to perform pulse phase-resolved polarization analysis of \fouru, we generated a timing model using the NICER data, which bracketed the \ixpe observation span (see Table \ref{tab:observations}). A simple Fourier power spectrum analysis of the NICER data showed pulsations at a period of $\sim7.668$~s. With this initial estimate, we generated 64 pulse times of arrival (TOAs) using the \texttt{photon\_toa.py} script in the NICERsoft\footnote{https://github.com/paulray/NICERsoft} data analysis repository. They were constructed using a Gaussian pulse template and imposing an integration time of 300~s (with a minimum exposure time of 200~s). The final timing parameters were then obtained through a weighted least-squares fit in the \texttt{PINT} pulsar data analysis package, by computing model corrections for the residuals between the initial TOAs and the pulse phase model involving the first frequency derivative. 
The derived pulse phase ephemeris with the NICER data is shown in Table~\ref{tab:ephemeris}. With the final timing model derived from the NICER data, we computed the pulse phases for \ixpe event times in the solar system barycenter reference frame. The folded 2.0--8.0 keV pulse profile is shown in Figure~\ref{fig:pulseprofile}.

\fouru exhibits strong energy dependence in the pulse profile over the 0.3--12 keV band, which is consistent with previous observations of the source by other missions \citep{beri-pulse-phase_2014,beri_pulse-phase-emission_2015}. However, for this work, we concentrate our analysis in the 2--8 keV band, over which the pulse profile does not exhibit significant changes. Detailed timing analysis of the \nicer data
is beyond the scope of this paper.

\begin{deluxetable*}{lr}
\tablecaption{Spin ephemeris of \fouru\ from NICER \label{tab:ephemeris}}
\tablewidth{0pt}
\tablehead{
Parameter & Value \\ 
}
\startdata
Spin Frequency, $\nu_0$ (Hz) & 0.1304042949(7) \\
Spin Frequency Derivative, $\dot\nu$ (Hz/s) & 2.33(7)$\times10^{-13}$ \\ 
Reference Phase Epoch, $t_0$ (TDB) & MJD 59662.72595304 \\ 
\tableline 
Terrestrial Time Standard (CLK) & TT (BIPM2019) \\ 
Solar System Ephemeris & JPL DE421 \\
Reference Epoch, TZRMJD (MJD) & 59656.0 \\
\enddata 
\end{deluxetable*}

\subsection{Pulse Phase-Resolved Polarization} \label{sec:polarization_dep}

For pulse phase-resolved polarization analysis, pulse phase intervals were identified using the \nicer pulse profile (Figure~\ref{fig:pulseprofile}) due to its high signal/noise.
Intervals 0.035--0.135 and 0.265--0.345 correspond to profile peaks that bracket a minimum in the phase range 0.135-0.265; the fourth interval spans the relatively flat pulse profile over the phase range 0.345--0.035 (modulo 1).
Physically, these intervals may have different origins that may relate to their polarizations.

\begin{figure}[t]
		\centering
		\includegraphics[width=\linewidth]{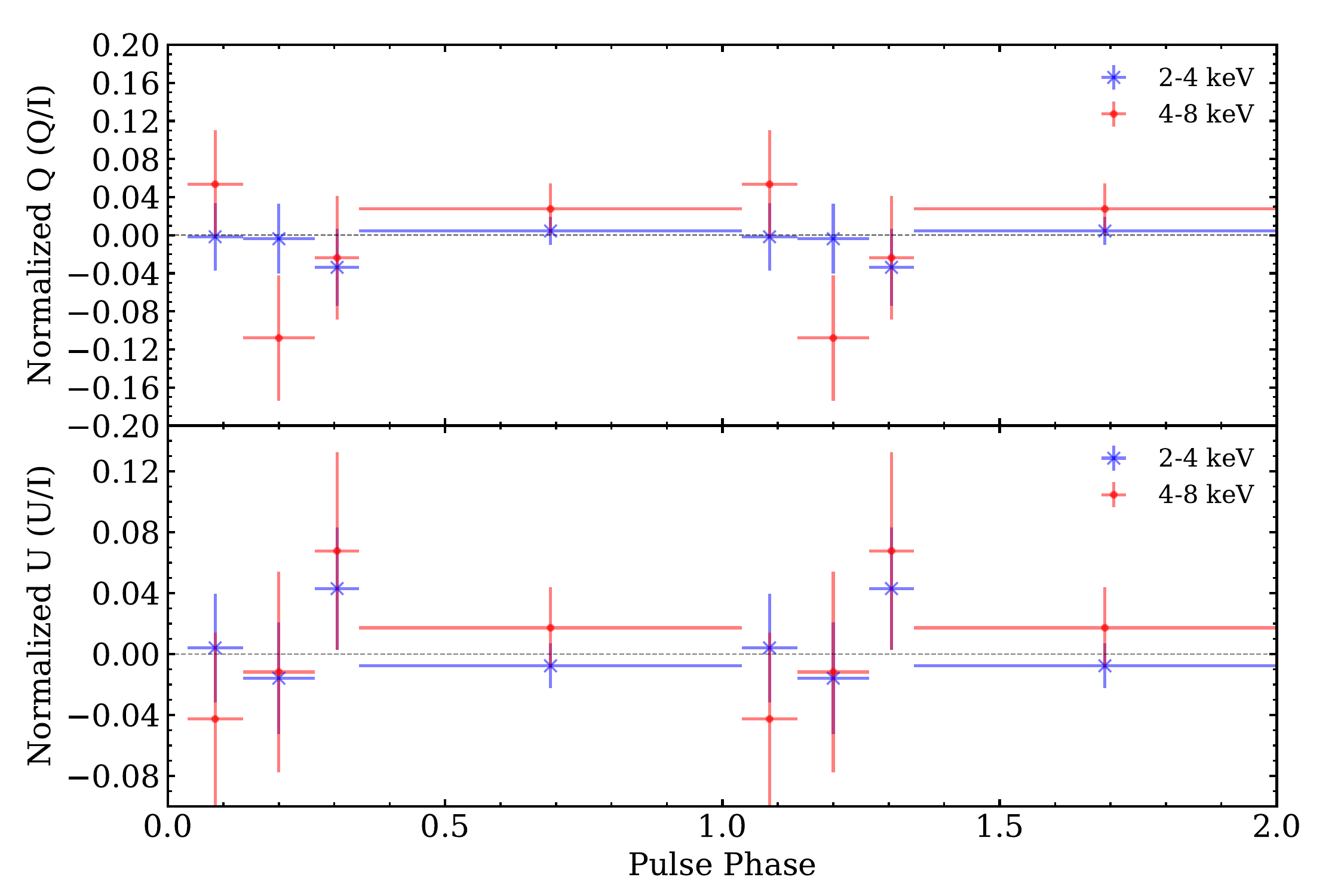}
		\caption{Normalized $Q$ ($Q/I$; top) and normalized $U$ ($U/I$; bottom) as a function of pulse phase in the 2--4 keV (blue) and 4--8 keV (red) energy bands. Two rotation cycles are plotted for clarity. The error bars denote $1\sigma$ uncertainties for one degree of freedom.  There is no clear detection of polarized emission in either energy band for any phase interval.}
		\label{fig:normu_normq}
	\end{figure}

The $q$ and $u$ values are given in Table~\ref{tab:qu_table} and shown in Figure~\ref{fig:normu_normq} for the four pulse phase intervals and in two energy bands, 2.0--4.0 keV and 4.0--8.0 keV.  Full band values are also given in Table~\ref{tab:qu_table}.
For the two bands, 10 of 16 (independent) $q$ and $u$ measurements both contain zero within their 1$\sigma$ error regions, which is consistent with random, unpolarized data.
Thus, we do not find any significant pulse phase-dependence in any energy band.



\begin{deluxetable}{cccccc}[htbp]
\caption{Normalized Stokes parameters in energy-phase intervals\tablenotemark{a}}
\tablehead{
\colhead{Energy} & \colhead{} & \multicolumn{4}{c}{Pulse Phase}  \\
\cline{3-6}
\colhead{(keV)} & \colhead{} & \colhead{0.035--0.135} & \colhead{0.135--0.265} & \colhead{0.265--0.345} & \colhead{0.345--0.035} 
}
\startdata
2.0--4.0  & $q$ & $-0.0016\pm0.0357$ & $-0.0037\pm0.0365$ & $-0.034\pm0.040$ & $0.0044\pm0.0147$ \\ 
             & $u$ & $0.0041\pm0.0357$ & $-0.016\pm0.037$ & $0.043\pm0.040$ & $-0.0077\pm0.0147$ \\
\hline 
4.0--8.0  & $q$ & $0.054\pm0.057$ & $-0.11\pm0.07$ & $-0.024\pm0.065$ & $0.028\pm0.027$ \\ 
             & $u$ & $-0.043\pm0.057$ & $-0.012\pm0.067$ & $0.068\pm0.065$ & $0.017\pm0.027$ \\ 
\hline 
2.0--8.0 & $q$ & $0.036\pm0.044$ & $-0.070\pm0.049$ & $-0.027\pm0.051$ & $0.019\pm0.020$ \\ 
             & $u$ & $-0.028\pm0.044$ & $-0.013\pm0.049$ & $0.060\pm0.051$ & $0.0081\pm0.0199$ \\
\hline 
\enddata
\tablenotetext{a}{The uncertainties are $1\sigma$ confidence limits for one degree of freedom.}
\label{tab:qu_table}   
\end{deluxetable}

\subsection{Spectropolarimetric Fitting} \label{sec:spectra}

\subsubsection{Broadband spectrum, phase-averaged} \label{sec:broadband}

 	\begin{figure}[htbp]
		\centering
		\includegraphics[width=.65\linewidth]{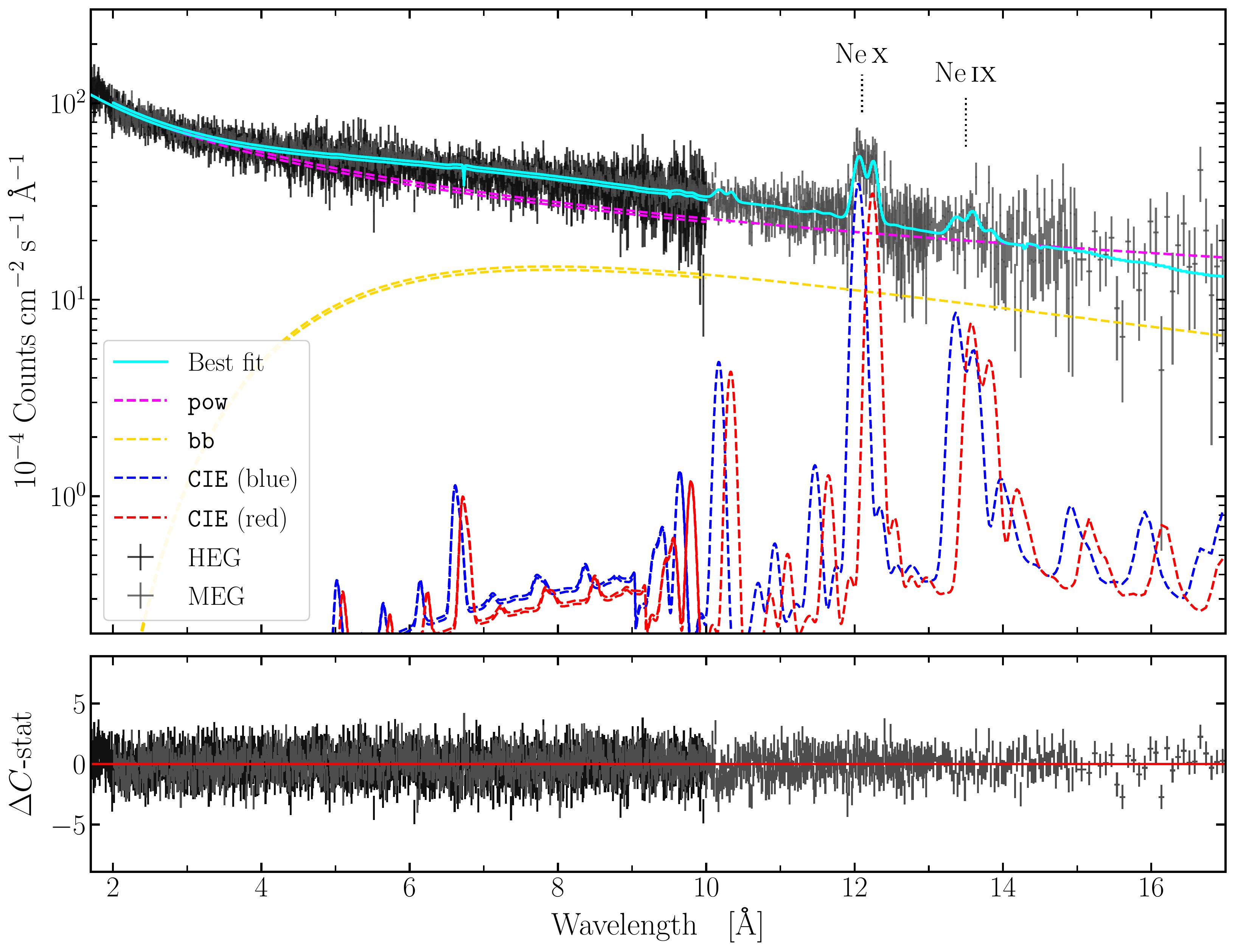}
		\caption{HETGS spectrum of \fouru. In the top-panel, the best-fit model of the HEG (in black) and MEG (in gray) spectra is represented with the cyan solid line. The different unabsorbed continuum components, the power-law and black body, are overplotted with dashed lines (magenta and yellow respectively). We also show the red- and blue-shifted collisional ionization models used to characterize the strong \ion{Ne}{9} and \ion{Ne}{10} lines. In the bottom panel we show the residuals of the best fit. } 
		\label{fig:hetg_spec}
	\end{figure}
	
The HETGS continuum spectra of \fouru are well fit using \texttt{spex} by a two-component model consisting of a power-law with a photon index of $\Gamma = 1.16\pm0.02$ and a black-body with a temperature of $k{T} = 0.40\pm 0.01\;\rm keV$. Both components are subject to absorption by neutral interstellar medium which we model using the \texttt{hot} component with a temperature fixed to $0.008\; \rm eV$ ($\sim 100\; \rm K$) in order to mimic the cold gas. 

The spectrum of \fouru shows the lines of H-like Ly$\alpha$ \ion{Ne}{10} and He-like \ion{Ne}{9}. High-resolution X-ray spectroscopic studies resolved the broad \ion{Ne}{10} feature into double-peak line, modeled as a collisionally ionized plasma located close to the inner-edge of the magnetically-truncated accretion \citep{Schulz19,Hemphill21}.
For the purposes of this paper, we used red- and blue-shifted \texttt{cie} (collisionally ionized plasma) components \citep{Kaastra96} to model the Keplerian line profile. We tied the plasma temperatures of the two
components and we let the Doppler shifts, turbulent velocities, and normalizations differ.

The HETGS spectrum and best-fit model are shown in Figure~\ref{fig:hetg_spec}. We list the parameter values in Table \ref{tab:hetg_fit} with the observed fluxes in the 0.5--2 and 2--10~keV energy bands. {The temperature of the collisionally ionized plasma, $T\simeq 8\times10^6 \rm\, K$, together with its turbulence and shift velocities are consistent with the previous \chandra observations \citep{Schulz19,Hemphill21}. We do not detect any significant secondary lower-temperature component. The \ovii triplet ($E\sim21.6\:\mbox{--}\:22.1\rm \; \text{\AA}$), which is used to constrain the cooler plasma, is outside the present HETGS energy band.} During the fit, we fixed the line-of-sight hydrogen column density to be consistent with the spectral analysis of the \nicer and \ixpe data (see \S~\ref{sec:phaseresolved}). Analysis of possible pulse phase dependence of the emission lines is beyond the scope of this paper. 
	
\begin{table}[htbp]
\centering 
\caption{Best-fit parameter values of the broadband HETGS spectra. 
}  
\begin{tabular}{cccc}
\hline \hline
Model & Parameter & Fit Value  & Units \\
\hline 
\multirow{2}{*}{\tt hot} & $N_{\rm H}$           & $0.14$           & $10^{22}{\rm\,cm^{-2}}$ \\ 
                         & $kT_{\rm ISM}$        & $0.008$          & eV                      \\ \cline{2-4}
\multirow{2}{*}{\tt pow} & $\Gamma$              & $1.15\pm0.2$     &  --                      \\
                         & $N_{\rm PL}$          & $0.38\pm0.01$    & $10^{44}\; \rm ph\;s^{-1}\;keV^{-1}$ \\ \cline{2-4}
\multirow{2}{*}{\tt bb}  & $kT_{\rm BB}$         & $0.406\pm0.005$  &  keV                    \\
                         & $N_{\rm BB}$          & $2.1\pm0.1$      & $10^{12}\rm \; cm^{-2}$ \\\cline{2-4}
\multirow{7}{*}{\tt cie} & $kT_{\rm CIE}$        & $ 0.69\pm0.08$        &  keV           \\
                         & $v_{\rm rms,\;red}$   & $ 1500^{+400}_{-600}$ &  km\,s$^{-1}$  \\
                         & $v_{\rm redshift}$    & $ 2700^{+300}_{-600}$ &  km\,s$^{-1}$  \\
                         & $N_{\rm red}$         & $ 1.0\pm0.6 $         &  $10^{68}\rm \; cm^{-3}$ \\
                         & $v_{\rm rms,\; blue}$ & $ 1600\pm300 $        &  km\,s$^{-1}$  \\
                         & $v_{\rm blueshift}$   & $ 2100\pm200 $        &  km\,s$^{-1}$  \\
                         & $N_{\rm blue}$        & $ 1.1\pm0.6$          & $10^{68}\rm \; cm^{-3}$ \\\cline{2-4}
\multirow{2}{*}{Flux}    & $F_{0.5-2\;\rm keV}$  & $ 2.6\pm0.1$          & $10^{-10}\rm \; erg\;s^{-1}\;cm^{-2}$ \\
                         & $F_{2-10\;\rm keV}$   & $ 0.69\pm0.05$        & $10^{-10}\rm \; erg\;s^{-1}\;cm^{-2}$ \\
\hline
\multicolumn{2}{c}{$C$-stat/d.o.f} & \multicolumn{2}{c}{$2930/2853$} \\ 
 \hline 
\end{tabular}
\label{tab:hetg_fit}   
\end{table}

The \nicer pulse phase-averaged spectrum over 0.4--10.0 keV is generally well described with an absorbed power law and blackbody, along with two Gaussians that represent the emission lines as found in the HETGS spectra. {The pulse phase-averaged values for the power-law index ($\Gamma=1.06$) and blackbody temperature ($kT = 0.37$~keV) are within the range of known values for the source from past Einstein, ASCA, Suzaku, and RXTE observations \citep{angelini_neon_1995,camero-arranz_4u_2012}. The broad emission lines around 0.6 keV and 1.0 keV likely correspond to a complex of oxygen lines and neon lines, respectively \citep{beri_pulse-phase-emission_2015}, as observed by XMM-Newton. Detailed comparisons of the spectral parameters relative to past behavior of the source is outside the direct scope of this work and will be reported in a future publication.}  Each of the two continuum components had a constant polarization model applied to them. We simultaneously fit the seven NICER spectra from each of the ObsIDs,\footnote{We note that for the spectrum corresponding to NICER ObsID 5661010203, the spectral fit was performed over 0.4--9.0 keV.} and then introduced the nine IXPE spectra ($I$, $Q$, and $U$ spectra for each of the three detectors). We fixed all of the spectral parameters except for the polarization degree and angle for each of the two continua, which were tied across all spectra. We also multiplied the model by a constant in order to account for cross-calibration uncertainties between the \nicer and \ixpe detectors, where we fix a value of 1 for the first \nicer observation.  In \texttt{XSPEC} parlance, the spectral model is \texttt{tbabs(powerlaw*polconst+bbodyrad*polconst+gauss+gauss)*const}.\footnote{We applied a 2\% systematic error for the spectral fitting.}

The results of the spectropolarimetric fit are given in Table \ref{tab:phaseave_spectralfit}, where uncertainties are given to $90\%$ confidence limits. The power-law index, blackbody temperature, and Gaussian centroid energies generally agree with that of previous studies of \fouru. We find that the polarization degrees of either of the two continua components are formally consistent with zero at 1$\sigma$.

\begin{deluxetable}{cc}[htbp]
\caption{Spectropolarimetric parameters from joint \ixpe/\nicer fits\tablenotemark{a}}
\tablehead{
\colhead{Parameter} & \colhead{Value}
}
\startdata
$\Gamma$ & $1.064_{-0.016}^{+0.011}$ \\
$N_{\rm PL}$ ($10^{-2}{\rm\,photons\,keV^{-1}}$ ${\rm\,cm^{-2}\,s^{-1}}$) & $1.98_{-0.05}^{+0.04}$ \\
$kT_{\rm BB}$ (keV) & $0.367_{-0.004}^{+0.005}$ \\ 
$N_{\rm BB}$\tablenotemark{\footnotesize b} & $243_{-11}^{+10}$ \\ 
$E_1$ (keV) & $0.636_{-0.006}^{+0.005}$ \\ 
$\sigma_1$ (keV) & $0.064_{-0.011}^{+0.018}$ \\ 
$N_{E_1}$ ($10^{-3} {\rm\ photons\,cm^{-2}}$ ${\rm\,s^{-1}}$) & $2.2_{-0.3}^{+0.6}$ \\
$E_2$ (keV) & $1.005_{-0.004}^{+0.004}$ \\ 
$\sigma_2$ (keV) & $0.032_{-0.007}^{+0.007}$ \\ 
$N_{E_2}$ ($10^{-3} {\rm\ photons\,cm^{-2}}$ ${\rm\,s^{-1}}$) & $1.33_{-0.11}^{+0.15}$ \\
$f_1$\tablenotemark{\footnotesize c} & 1 \\ 
$f_2$ & $1.050_{-0.007}^{+0.007}$ \\
$f_3$ & $1.106_{-0.007}^{+0.007}$ \\
$f_4$ & $1.075_{-0.006}^{+0.006}$ \\ 
$f_5$ & $1.143_{-0.009}^{+0.009}$ \\ 
$f_6$ & $1.042_{-0.007}^{+0.007}$ \\ 
$f_7$ & $1.169_{-0.007}^{+0.007}$ \\
$f_{\rm DU1}$ & $0.993_{-0.008}^{+0.008}$ \\
$f_{\rm DU2}$ & $1.013_{-0.008}^{+0.008}$ \\ 
$f_{\rm DU3}$ & $0.854_{-0.007}^{+0.007}$ \\ 
\hline 
$F_{2-10}{\rm\,(10^{-10}\, erg\,s^{-1}\,cm^{-2})}$\tablenotemark{\footnotesize d} & $2.36_{-0.03}^{+0.03}$ \\ 
PD$_{\rm PL}$ & $0.023_{-0.023}^{+0.024}$ \\ 
PA$_{\rm PL}$ (deg.) & \nodata\tablenotemark{\footnotesize e} \\
PD$_{\rm BB}$ & $0.18_{-0.18}^{+0.18}$ \\ 
PA$_{\rm BB}$ (deg.) & \nodata\tablenotemark{\footnotesize e} \\
$\chi^2$ (d.o.f.) & $2809.75\ (2303)$ \\
\hline 
\enddata
\tablenotetext{a}{The uncertainties quoted are given with 90\% confidence limits. The column density is fixed at $1.4\times10^{21}{\rm\,cm^{-2}}$.}
\tablenotetext{b}{In units of $R_{\rm km}^2/D_{10}^2$, with $R_{\rm km}$ being the radius of the emitting region, and $D_{10}$ is the distance to \fouru\ in units of 10 kpc.}
\tablenotetext{c}{The multiplicative constant parameterizes the cross-calibration uncertainties between the different \nicer observations as well as the between the \ixpe detector units. See text for details.}
\tablenotetext{d}{2--10 keV absorbed flux corresponding to the spectrum from the first \nicer observation.}
\tablenotetext{e}{When the polarization degree is consistent with zero at 90\% confidence, the polarization angle range is unconstrained.}
\label{tab:phaseave_spectralfit}   
\end{deluxetable}

\subsubsection{Pulse phase-resolved} \label{sec:phaseresolved} 

We also conducted pulse phase-resolved spectropolarimetric fits with the same pulse phase intervals that were defined in Section \ref{sec:polarization_dep}. The spectral fitting process was conducted the same way as that of the pulse phase-averaged case.
The polarization degree and angle were allowed to vary with phase, though they were tied across data sets.
The results from the pulse phase-resolved spectropolarimetric analysis are presented in Table \ref{tab:phasedep_spectralfit}, with $90\%$ confidence intervals reported for the uncertainties.

Consistent with previous work, we find that the parameters of the spectral model vary with pulse phase.  {As with the discussion on the pulse phase-averaged spectrum, we leave detailed discussion of any variation in the spectral parameters to future work.}
More important for the purposes of this paper are the polarization parameters.
The polarization degree of the blackbody component is formally consistent with zero for all four pulse phase intervals; in such cases, the error range on the EVPA is indeterminate.
By contrast, the 90\% confidence bands for the polarization degree of the power-law component are not consistent with zero in two out of four intervals. To examine the significance of these polarization measurements further, we computed the probability that the polarization degree is zero, $P(>\Delta \chi^2)$, for each interval. We find that $P(>\Delta \chi^2) \lae 0.38$ for each interval and is below 0.1 for two intervals. The probability that randomly polarized data would give probabilities at 0.38 or less for all four observations is less than 2\%.  Other tests of the set of probabilities against the expectations of a uniform distribution, such as the Kolmogorov-Smirnov and Anderson-Darling tests, yielded similar results.  {This collection of four probabilities is equivalent to slightly better than a 2 $\sigma$ result; while it is a statistically unlikely event, it must be considered to be marginal evidence.}
Nevertheless, the result does indicate that there {may be} a small average polarization that is not detectable when pulse phase averaged due to EVPA swings, which are indicated in Table~\ref{tab:phasedep_spectralfit}.
An additional analysis in which PD$_{\rm PL}$ was tied across phase bins gave a polarization fraction of $4.8 \pm 2.3$\% (90\% confidence interval).

\begin{deluxetable}{ccccc}[htbp]
\caption{Phase-Dependent Spectropolarimetric Fit Results\tablenotemark{a}}
\tablehead{
\colhead{} & \multicolumn{4}{c}{Pulse Phase} \\
\cline{2-5}
\colhead{Parameter} & \colhead{0.035--0.135} & \colhead{0.135--0.265} & \colhead{0.265--0.345} & \colhead{0.345--1.035} 
}
\startdata
$\Gamma$ & $0.954_{-0.018}^{+0.018}$ & $1.19_{-0.02}^{+0.02}$ & $0.969_{-0.019}^{+0.019}$ & $1.00_{-0.02}^{+0.05}$ \\
$N_{\rm PL}$ & $2.29_{-0.06}^{+0.06}$ & $1.93_{-0.07}^{+0.06}$ & $2.25_{-0.06}^{+0.06}$ & $1.73_{-0.7}^{+0.16}$  \\
$kT$ (keV) & $0.325_{-0.008}^{+0.009}$ & $0.377_{-0.009}^{+0.009}$ & $0.317_{-0.006}^{+0.007}$ & $0.380_{-0.012}^{+0.005}$ \\ 
$N_{\rm BB}$ & $223_{-26}^{+26}$ & $146_{-15}^{+16}$ & $301_{-31}^{+31}$ & $262_{-9}^{+18}$ \\ 
$E_1$ (keV) & $0.646_{-0.008}^{+0.008}$ & $0.649_{-0.007}^{+0.007}$ & $0.652_{-0.008}^{+0.008}$ & $0.603_{-0.013}^{+0.028}$ \\ 
$\sigma_1$ (keV) & $0.057_{-0.012}^{+0.013}$ & $0.074_{-0.011}^{+0.012}$ & $0.033_{-0.014}^{+0.013}$ & $0.116_{-0.050}^{+0.018}$ \\ 
$N_{E_1}$ & $2.1_{-0.4}^{+0.4}$ & $3.5_{-0.5}^{+0.6}$ & $1.5_{-0.3}^{+0.3}$ & $3.9_{-1.9}^{+0.9}$ \\
$E_2$ (keV) & $1.011_{-0.005}^{+0.005}$ & $0.989_{-0.006}^{+0.006}$ & $1.004_{-0.008}^{+0.007}$ & $1.004_{-0.003}^{+0.004}$ \\ 
$\sigma_2$ (keV) & $0.033_{-0.010}^{+0.009}$ & $0.067_{-0.008}^{+0.008}$ & $0.043_{-0.013}^{+0.013}$ & $0.034_{-0.012}^{+0.006}$ \\ 
$N_{E_2}$ & $1.32_{-0.15}^{+0.17}$ & $2.3_{-0.2}^{+0.2}$ & $1.5_{-0.2}^{+0.2}$ & $1.43_{-0.26}^{+0.12}$ \\
$f_1$\tablenotemark{b} & - & - & - & 1 \\ 
$f_2$ & - & - & - & $1.053_{-0.005}^{+0.005}$ \\
$f_3$ & - & - & - & $1.104_{-0.006}^{+0.006}$ \\
$f_4$ & - & - & - & $1.075_{-0.005}^{+0.005}$ \\ 
$f_5$ & - & - & - & $1.139_{-0.008}^{+0.008}$ \\ 
$f_6$ & - & - & - & $1.065_{-0.006}^{+0.006}$ \\ 
$f_7$ & - & - & - & $1.167_{-0.006}^{+0.006}$ \\
$f_{\rm DU1}$ & - & - & - & $0.977_{-0.008}^{+0.008}$ \\
$f_{\rm DU2}$ & - & - & - & $0.992_{-0.008}^{+0.008}$ \\ 
$f_{\rm DU3}$ & - & - & - & $0.843_{-0.007}^{+0.007}$ \\ 
\hline 
$F_{2-10}$ & $3.19_{-0.03}^{+0.03}$ & $1.86_{-0.019}^{+0.018}$ & $3.06_{-0.04}^{+0.03}$ & $2.316_{-0.015}^{+0.016}$ \\ 
PD$_{\rm PL}$ & $0.046_{-0.046}^{+0.048}$ & $0.10_{-0.07}^{+0.06}$ & $0.10_{-0.07}^{+0.06}$ & $0.03_{-0.03}^{+0.03}$ \\ 
PA$_{\rm PL}$ (deg.) & $-17$\tablenotemark{e} & $-19_{-23}^{+23}$ & $53_{-21}^{+22}$ & $-32$\tablenotemark{e} \\
$P(>\Delta \chi^2)$\tablenotemark{c} & 0.377 & 0.080 & 0.059 & 0.265 \\
PD$_{\rm BB}$\tablenotemark{d} & \nodata & \nodata & \nodata & \nodata \\ 
PA$_{\rm BB}$\tablenotemark{e} (deg.) & \nodata & \nodata & \nodata & \nodata \\ 
$\chi^2$ (d.o.f.) & \multicolumn{4}{c}{8244.07 (7573)} \\
\hline 
\enddata
\tablenotetext{a}{Parameters, units, and notes are the same as in Table~\ref{tab:phaseave_spectralfit}.} 
\tablenotetext{b}{Cross-normalization constants are tied across the four phase bins.}
\tablenotetext{c}{$P(>\Delta \chi^2)$ is the probability that PD$_{\rm PL} = 0$ based on $\Delta \chi^2$ from the best fit with two degrees of freedom.}
\tablenotetext{d}{PD$_{\rm BB}$ is consistent with all values from zero to one
at 90\% confidence for all pulse phase intervals.}
\tablenotetext{e}{When PD is consistent with zero at 90\% confidence, the polarization angle range is unconstrained. }
\label{tab:phasedep_spectralfit}   
\end{deluxetable}

\subsection{Rotating Vector Model Fits}
\label{sec:rvm}

It is natural to expect the orientation of the polarization to change with the phase of the pulsar; therefore, we can impose a model on the expected evolution of the polarization with phase to understand the pulsar further.   Although the magnetic field near the star and near the accretion disk is probably complicated, in between it is expected to be approximately dipolar. 
The polarized radiation propagates through the magnetosphere in the normal modes perpendicular and parallel to the local magnetic field until reaching the polarization-limiting radius, which is much larger than the star and much smaller than the inner edge of the accretion disk \citep[e.g.,][]{2018Galax...6...76H}. 
In this situation, the rotating-vector model (RVM) of \citet{RC69} holds to a good approximation.
In the RVM, the PA (measured from north to east) is given by \citep{Poutanen20RVM,herx1_ixpe} 
\begin{equation} \label{eq:pa_rvm}
\tan (\mbox{PA}-\chi)= \frac{\sin \theta\ \sin (\phi-\phi_0)}
{  \cos i \sin \theta  \cos (\phi-\phi_0) -\sin i \cos \theta } ,
\end{equation}
where $\chi$ is the position angle of the pulsar spin, $i$ is the pulsar inclination (i.e., the angle between the pulsar spin vector and direction to an observer), $\theta$ is the magnetic obliquity (i.e., the angle between the magnetic dipole axis and the spin axis), $\phi_0$ is the phase when the magnetic axis is closest to the observer (the right ascension of the observer), and $\phi$ is the pulse phase.
  
\begin{figure*}
    \centering
    \includegraphics[width=\linewidth]{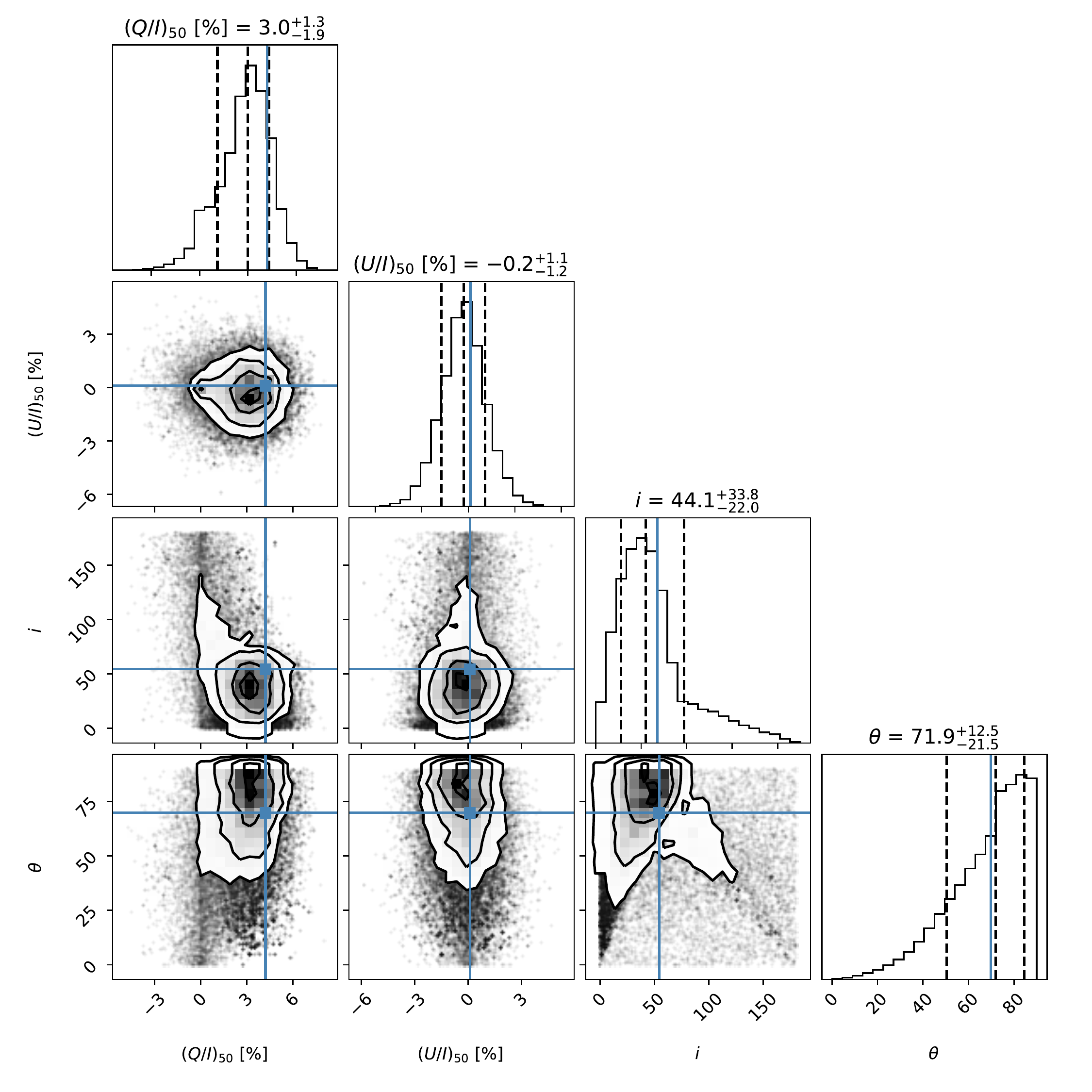}
    \caption{Weighted MCMC with a prior on $i$ and $\theta$ that is uniform in solid angle.  The two RVM parameters that represent the position angle ($\chi$) of the spin axis and the right ascension of the observer ($\phi_0$) are not well constrained individually but their sum is, which is represented as the Stokes parameters at a phase of 50\%.  The blue lines indicate the maximum likelihood without a prior on the geometry.  The values and uncertainties are the 16th, 50th and 84th percentiles of the one-dimensional posterior distributions.  The contours contain 39.3\%, 67.5\% and 86.5\% of the distribution in two dimensions. }
    \label{fig:emcee_binned}
\end{figure*}

We fit the RVM to the measured photo-electron azimuthal angle using an unbinned likelihood method as outlined by \citet{unbinned} and \citet{2021AJ....162..134M}.
As the RVM is fit to the individual photon arrivals, it is independent of the previous binned results.
Unfortunately, there are too few photons to constrain the geometry of the system strongly, as shown in Figure~\ref{fig:emcee_binned}. 
In the rotating frame, the fits yield a mean PD of $3.2_{-1.5}^{+1.2}\%$ when marginalized over the geometric parameters.  Because the PA apparently spins around in the sky, it is difficult to constrain the position angle of the spin axis $\chi$ and the right ascension of the observer $\phi_0$; however, the polarization at a particular phase (50\%) is better constrained and points approximately North. From the posterior distribution of $(Q/I)_{50}$, we find that 5.5\% of the distribution has $(Q/I)_{50}<0$, so we argue that the polarization of the source is weakly detected at the 94\%-confidence level (about $1.9\sigma$), consistent with the results from phase-dependent spectropolarimetry
(\S \ref{sec:phaseresolved}).

\section{Summary and Discussion} \label{sec:discussion}

We do not find any significant {\em phase-averaged} polarization of \fouru.
By contrast,
the pulse-phased spectropolarimetric modeling provided marginal evidence that the power-law component may be
polarized at the 4.8$\pm$2.3\% level, with an EVPA that changes with pulse phase.
The RVM fits indicate polarization of about 3\%, significant at the 2$\sigma$ level, with
modest inclination ($i$ in the range of 20\deg--80\deg) and relatively high
magnetic obliquity (50\deg--85\deg).
However, these measurements are highly uncertain;
more exposure will be necessary for detailed modeling of the pulse phase dependent polarization of \fouru.
{For Her X-1 and Cen X-3, the pulse-phased X-ray polarization averaged 7-10\%, much lower than many of the predictions.  These two sources yielded sufficient signal to obtain significant polarization detections \citep{herx1_ixpe,2022arXiv220902447T}.  While \fouru was fainter than these two pulsars, the low polarization is consistent with the general finding that X-ray polarizations are lower than expected.}

{Before the launch of \ixpe, we expected that} the X-ray emission from magnetized $\gtrsim 10^{12}{\rm\,G}$ neutron stars {would} be polarized, due to the strong dependence of the two modes on the photon mean free path. In particular, this {expectation} is due to 
{a large difference in the  mean free path of extraordinary (X) mode photons compared to the ordinary (O) mode photons}
\citep{Lodenquai1974,Gnedin1974,Pavlov2000, Potekhin2003}.
The effect is most important for energies observable by \ixpe, which are below the electron cyclotron energy: 37 keV for \fouru.
Generally, the pencil beam model gives lower polarizations at energies in the middle of the \ixpe band than the fan beam for a wide range of $i$ and $\theta$
\citep{meszaros88}. \citet{meszaros88} show at least one case ({$E = 3.8$ keV}, $i= 50$\deg, and $\theta = 20$\deg) 
where the {phase-dependent polarization fraction does not appear to exceed $5$\%, consistent with our results for} \fouru.
These geometric parameters are similar to those obtained by
\citet{kii86}, who generated pulse light curve models over the 0.6--30 keV band that approximately match the observations for $i=40$\deg{} and $\theta = 50$\deg.
\citet{kii87} also found that the polarization can be quite small in the energy band well below the CRSF  if the $\theta$ is below 30\deg.

For \fouru, the posterior probability plots from the weighted MCMC analysis in Figure~\ref{fig:emcee_binned} imply a preference for larger values of $\theta$. We caution against over-interpretation given the large uncertainties (and low number of source photons), but this indication is intriguing. The higher values of $\theta$ might indicate that a single-geometry picture is insufficient to characterize the accretion geometry around the pulsar.
The \ixpe observations of Her X-1 support a two component model \citep{herx1_ixpe}.
In fact, the complex nature of the accretion geometry around accreting pulsars has been suggested for a long time. For example, \cite{1996ApJ...467..794K} analyzed energy-dependent pulse profiles of Cen X-3 and modeled the pulse profiles by decomposing them into two single-pole
pulse profiles, i.e., two emission regions, and suggested that the observed pulse profile can be explained by a combined fan-beam and pencil-beam geometry.  {This decomposition is supported by \ixpe observations of Cen X-3, another subcritical accretor \citep{2022arXiv220902447T}.} Similarly, \cite{2008A&A...491..833K} looked at pulse phase-resolved spectroscopy of EXO 2030+375 and showed that the temporal evolution of the pulse profiles is qualitatively consistent with the transition from a fan-beam geometry at the outburst peak to the combination of fan-beam and pencil-beam geometries towards the tail of the outburst decay.

Modeling of the accretion column in accreting X-ray pulsars by \cite{2012A&A...544A.123B} suggests that the emergence of the accretion geometry is a luminosity-dependent behavior. They defined a critical luminosity $L_{\rm crit} = 1.5\times10^{37} B_{12}^{16/15}{\rm\,erg\,s^{-1}}$, where $B_{12}$ is the surface magnetic field strength in units of $10^{12}$ G. For $L < L_{\rm crit}$, the thermal emission escapes out of the top of the accretion hot spot; for $L > L_{\rm crit}$, a radiation-dominated shock forms above the surface which results in radiation escaping through the walls of the accretion column. Under this model, there is an intermediate range of luminosities ($L_X \lesssim 10^{35-37}{\rm\,erg\,s^{-1}}$) where the accretion geometry can be described by a hybrid fan-beam plus pencil-beam geometry \citep{2000ApJ...529..968B,2012A&A...544A.123B}.  

The luminosity-dependent behavior of evolving accretion geometry has been suggested in the past for observations of the pulse profiles of the pulsars in Be X-ray binaries EXO 2030+375 \citep{1989ApJ...338..373P} and GX 304--1 \citep{2015A&A...581A.121M}. For \fouru, a 37 keV CRSF implies a field strength of $B \sim (3-4)\times10^{12}{\rm\,G}$ \citep{orlandini_bepposax_1998}, which suggests $L_{\rm crit}\sim(4.8-6.6)\times10^{37}{\rm\,erg\,s^{-1}}$. Assuming a distance of 3.5 kpc \citep{Schulz19}, the 0.3--100 keV luminosity of the source from the latest observations (see Table \ref{tab:phaseave_spectralfit}; extrapolation was done with \texttt{energies extend} in \texttt{XSPEC}) is $L \approx 3.7\times10^{36}{\rm\,erg\,s^{-1}}$, which puts \fouru in the intermediate luminosity range of accreting X-ray pulsars that could have a hybrid accretion geometry \citep{2012A&A...544A.123B}. \fouru is known to vary in luminosity over long timescales, and it is surmised that the accretion torque is correlated with the luminosity \citep{camero-arranz_new_2010}. \fouru underwent a torque reversal from a sustained spin-down to sustained spin-up state some time in 2008 \citep{camero-arranz_new_2010}. Suzaku observations in 2006 and 2010 showed that the source luminosity (0.5--10 keV) had increased from $3\times10^{35}{\rm\,erg\,s^{-1}}$ to $7.4\times10^{35}{\rm\,erg\,s^{-1}}$ \citep{camero-arranz_4u_2012}, still well within the intermediate luminosity regime which suggests a hybrid accretion geometry.

\fouru is known to have strongly energy-dependent pulse profiles that differ in morphology in the spin-up and spin-down states \citep{beri-pulse-phase_2014}. \cite{iwakiri19} modeled the energy-dependence of the pulse profile by using a relativistic ray tracing code and found that the complex morphology of the pulse profiles can be explained by a hybrid fan-beam and pencil-beam pattern, where at lower energies ($\lesssim 10$ keV), both beam patterns contribute at similar levels to the pulse profile, and at higher energies ($\gtrsim 10$ keV), the pencil-beam pattern dominates the emission. They noted qualitative similarities to the work done by \cite{kii86}, and drew connections between their fan-beam (pencil-beam) emission and the O mode (X mode) emission in \cite{kii86}. The pulse phase-resolved spectropolarimetry with NICER and IXPE data that we showed here (see Table \ref{tab:phasedep_spectralfit}) suggests a swing in the polarization angle of $\sim90\deg$ across the trough and the second peak of the pulse profile, which would be consistent with a change in the dominance of the O and X mode emission \citep{kii86,iwakiri19}. Future IXPE observations at a higher flux state of \fouru will provide us with enough statistics to ascertain the accretion geometry at play here.

The observation campaign involving \ixpe, \nicer, and \chandra shows the value of using complementary instruments to understand the behavior of the source. \nicer provided an accurate timing solution as well as anchored the spectral behavior below 2.0 keV, where it has excellent soft response. The \chandra HETGS corroborates the spectral continuum determined by NICER.  A robust spectral model founded upon data from well studied instruments provides a solid foundation for a spectropolarimetric characterization of \fouru.  Only with a secure spectral model was it possible to make maximum use of \ixpe data, enabling detection of polarized flux.

\begin{acknowledgments} 

The Imaging X-ray Polarimetry Explorer (IXPE) is a joint US and Italian mission.  The US contribution is supported by the National Aeronautics and Space Administration (NASA) and led and managed by its Marshall Space Flight Center (MSFC), with industry partner Ball Aerospace (contract NNM15AA18C).  The Italian contribution is supported by the Italian Space Agency (Agenzia Spaziale Italiana, ASI) through contract ASI-OHBI-2017-12-I.0, agreements ASI-INAF-2017-12-H0 and ASI-INFN-2017.13-H0, and its Space Science Data Center (SSDC) with agreements ASI-INAF-2022-14-HH.0 and ASI-INFN 2021-43-HH.0, and by the Istituto Nazionale di Astrofisica (INAF) and the Istituto Nazionale di Fisica Nucleare (INFN) in Italy.  This research used data products provided by the IXPE Team (MSFC, SSDC, INAF, and INFN) and distributed with additional software tools by the High-Energy Astrophysics Science Archive Research Center (HEASARC), at NASA Goddard Space Flight Center (GSFC).  Funding for this work was provided in part by contract 80MSFC17C0012 from the MSFC to MIT in support of the \ixpe project.  Support for this work was provided in part by the National Aeronautics and Space Administration (NASA) through the Smithsonian Astrophysical Observatory (SAO)
contract SV3-73016 to MIT for support of the Chandra X-Ray Center (CXC),
which is operated by SAO for and on behalf of NASA under contract NAS8-03060.
MN also acknowledges support from the NASA NICER program under detector team grant 80NSSC19K1287 and guest observer grant 80NSSC22K1350. SST and JP were supported by the Russian Science Foundation grant 20-12-00364 and the Academy of Finland grants 333112, 349144, 349373, and 349906.
We thank K.\ C.\ Gendreau and Z.\ Arzoumanian for their help in arranging coordinated NICER scheduling of the observations.

\end{acknowledgments} 

\facilities{{\chandra(HETGS), \ixpe, \nicer}}
\software{{SPEX \citep{Kaastra20}, ixpeobssim \citep{Baldini2022}, PINT \citep{luo21}, Astropy \citep{astropy:2013, astropy:2018}, NumPy and SciPy \citep{virtanen20}, Matplotlib \citep{hunter07}, IPython \citep{perez07}, tqdm \citep{dacostaluis22}, HEASoft 6.30.1\footnote{http://heasarc.gsfc.nasa.gov/ftools} \citep{heasoft}}}

\bibliography{nstar_polarimetry,refs_4u1626,misc}
\bibliographystyle{aasjournal}



\end{document}